\documentclass{PoS}

\usepackage[utf8]{inputenc}
\usepackage{mathtools}
\usepackage{amsfonts} 
\usepackage{amssymb} 
\usepackage{amsmath} 
\usepackage{verbatim}

\newcommand{\Choose}[2]{{\begin{pmatrix} {#1} \\[-3pt] {#2} \end{pmatrix}}}
\newcommand{\Kdf}[0]{\mathcal K_{\mathrm{df},3}}

\newcommand{\Mdf}[0]{\mathcal M_{\mathrm{df},3}}
\newcommand{\Mth}[0]{\mathcal M_{ 3}}
\newcommand{\Mthr}[0]{\mathcal M_{ 3,\mathrm{th}}}

\newcommand{\K}[0]{\mathcal K_2}
\newcommand{\M}[0]{\mathcal M_2}
\newcommand{\PV}[0]{\widetilde{\mathrm{PV}}}

\title{Extracting three-body observables from finite-volume quantities}

\ShortTitle{Three-body observables from the lattice}

\author{\speaker{Maxwell T. Hansen}\\
        Institut f\"ur Kernphysik and Helmholz Institute Mainz, Johannes Gutenberg-Universit\"at Mainz,
55099 Mainz, Germany\\
        E-mail: \email{hansen@kph.uni-mainz.de}}

\abstract{Scattering and transition amplitudes with three-hadron final states play an important role in nuclear and particle physics. However, predicting such quantities using numerical Lattice QCD is very difficult, in part because of the effects of Euclidean time and finite volume. In this review we highlight recent formal developments that work towards overcoming these issues. We organize the presentation into three parts: large volume expansions, non-relativistic nonperturbative analyses, and nonperturbative studies based in relativistic field theory. 

In the first part we discuss results for ground state energies and matrix elements given by expanding in inverse box length, $1/L$. We describe complications that arise at $\mathcal O(1/L^6)$ and include a table summarizing the results of different calculations. 

In the second part we summarize three recent non-relativistic non-perturbative studies and highlight the main conclusions of these works. This includes demonstrating that the three-particle finite-volume spectrum is determined, up to exponentially suppressed effects, by on-shell amplitudes, as well as recovering a finite-volume quantization condition for scattering a stable particle off a two-particle bound state. In this part we also highlight recent work concerning a three-particle bound state in a finite volume. 

In the third and final part, we review recent work based in non-perturbative relativistic field theory. Here the finite-volume spectrum has been related to an intermediate infinite-volume quantity which itself is related via a known integral equation to the relativistic, model-independent three-particle scattering amplitude. We motivate the appearance of the intermediate quantity, explain how it is related to the standard amplitude, and discuss prospects for using the result to constrain three-particle observables.}

\FullConference{The 33rd International Symposium on Lattice Field Theory\\
		14 -18 July 2015\\
		Kobe International Conference Center, Kobe, Japan*}

\begin{document}


Over forty years after its formulation, we still have an incomplete understanding as to how quantum chromodynamics (QCD) reproduces the resonances observed in nature. There is overwhelming evidence that the observed spectrum arises from the fact that the QCD coupling grows as the energy scale is lowered. This results in quark confinement, so that the low-energy degrees of freedom are hadrons, built from quark constituents. The growth of the coupling also invalidates perturbation theory, necessitating the use of a non-perturbative approach, such as lattice QCD (LQCD), to study such states. 

To describe resonances one ideally aims to extract scattering amplitudes.
Continuing such amplitudes to poles in the complex energy plane then gives model-independent resonant masses and widths. However, in numerical LQCD it is only possible to access correlators with Euclidean time coordinates, at a finite lattice spacing, and in a finite volume. For hadronic scattering, the restriction to Euclidean time is particularly problematic. At large time separations the relevant matrix elements are dominated by off-shell states \cite{Maiani:1990ca} and analytic continuation of the numerical results to Minkowski-time correlators is generally an ill-posed problem \cite{Bertero}.

One can overcome this issue by viewing finite volume as a tool rather than an unwanted artifact. In particular, L\"uscher showed that for identical scalar particles the finite-volume energy spectrum is related to the two-to-two scattering amplitude~\cite{Luscher:1986n2,Luscher:1991n1}. This result has since been generalized to accommodate non-zero total momentum \cite{Rummukainen:1995vs,KSS,Christ:2005}, non-identical, non-degernate particles~\cite{Fu:2011xz,Leskovec:2012gb}, multiple, strongly-coupled two-particle channels~\cite{Liu:2005kr, Bernard:2010fp,Hansen:2012tf,Briceno:2012yi} and particles with intrinsic spin~\cite{Li:2012bi,Gockeler:2012yj,Briceno:2014oea, Briceno:2015csa}.\footnote{For numerical applications see, for example, Refs.~\cite{Dudek:2012xn,Pelissier:2012pi,Wilson:2014cna,Lang:2014yfa,Wilson:2015dqa,Bulava:2015qjz,Nicholson:2015pys,Detmold:2015qwf,Lang:2015sba,Orginos:2015aya,Junnarkar:2015jyf}.} 

The relations generally take the form of a quantization condition
\vspace{-2pt}
\begin{equation}
\label{eq:genqc}
\mathrm{det}[(\mathcal M_2)^{-1} + F] = 0 \,.
\vspace{-2pt}
\end{equation}
The energies for which this equation holds define the finite-volume spectrum of the theory. Here $\mathcal M_2$ is the two-to-two scattering amplitude, viewed as a matrix for a given fixed value of center of mass (CM) frame energy. In the case of a single channel of scalar particles $\mathcal M_2$ has indices $\ell, m; \ell', m'$ which denote the incoming and outgoing two-particle angular momentum. To incorporate multiple channels as well as nonzero spin one must enlarge this space to describe the additional degrees of freedom. $F$ is a matrix on the same space as $\mathcal M_2$ that encodes the effects of finite-volume. It is a known function, independent of the dynamics of the system.

Various features are common to the results of Refs.~\cite{Luscher:1986n2}-\cite{Briceno:2015csa} that are summarized by Eq.~(\ref{eq:genqc}). All results are valid up to neglected corrections of the form $e^{- m_{\pi} L}$, with $m_{\pi}$ the physical pion mass and $L$ the extent of the finite-volume. In addition, for all variations of the quantization condition, the matrix appearing inside the determinant is formally infinite dimensional. Thus, to use the relation in order to extract amplitudes, the matrices must be truncated. It turns out to be enough to truncate $\mathcal M_2$; one may then project $F$ to the same subspace with no additional approximation. Neglecting higher components of $\mathcal M_2$ at low energies is well motivated because these are suppressed by $p^{2 \ell}$ where $p$ is the magnitude of momentum in the CM frame and $\ell$ is the orbital angular momentum. In the case of identical scalar particles with zero total momentum, truncating to the s-wave gives
\begin{equation}
\label{eq:sqc}
  p \cot \delta  = \frac{1}{ \pi L}  \lim_{\Lambda \rightarrow \infty} \left[  \sum_{\vec n}^\Lambda   \frac{1}{  \vec n^2 -   [pL/(2\pi)]^2 } - 4 \pi \Lambda  \right ] \,,
\end{equation}
where $\delta$ is the s-wave scattering phase shift. For each finite-volume energy level, $E_n= 2 \sqrt{p_n^2 + m^2}$, in a given box size $L$, Eq.~(\ref{eq:sqc}) gives the s-wave scattering phase shift at $p_n$.

Also common to these formulations is the restriction to two-particle states. More precisely, the quantization conditions may be used up to the production threshold for more than two particles. Any resonance decaying into more than two hadrons is thus inaccessible using these results.

In this review we summarize progress towards relating finite-volume quantities and scattering observables in the three-particle sector. Such work is motivated in part by a wide variety of interesting three-particle resonances. For example the mass of the Roper resonance turns out to be lower than that of the negative-parity ground state, inconsistent with quark model predictions. Since the Roper decays with $40 \%$ branching fraction into $N \pi \pi$~\cite{Beringer:1900zz}, the lattice can only resolve the nature of this state with a reliable three-particle formalism. Beyond this, such formalism is a necessary first step towards describing any scattering or decay processes involving more than two hadrons. 

The remainder of this review is organized as follows: In the following section we describe results for energies and matrix elements, obtained by expanding in inverse box length, $1/L$. In Sec.~\ref{sec:nonrel} we summarize various non-perturbative studies based in non-relativistic field theory. Then, in Sec.~\ref{sec:rel} we describe work based in relativistic field theory that, with certain restrictions, relates the three-particle finite-volume spectrum with scattering amplitudes. We briefly conclude in Sec.~\ref{sec:conc}.

\section{Large volume expansions}

\label{sec:Lexp}

If the box length, $L$, is much larger than any other length scale in the system, then it is useful to express the ground state energy as a power series in $1/L$. This was already accomplished through $\mathcal O(1/L^5)$ by Huang and Yang in 1957 \cite{Huang}. In the notation of Ref.~\cite{Beane2007}, their result for the $n$-boson ground state in a periodic finite volume is 
\begin{multline}
  E^{(n)}(L) = n m +
  \frac{4 \pi a}{m L^3}\Bigg\{\Choose{n}{2}
  -\left(\frac{a}{\pi L}\right)\Choose{n}{2} {\cal I} \\
+\left(\frac{a}{\pi L}\right)^2 \left\{
\Choose{n}{2}{\cal I}^2
-\left[\Choose{n}{2}^2 -12\Choose{n}{3}-6\Choose{n}{4}\right]{\cal J}
\right\} \Bigg\} +{\cal O}\left(1/L^{6}\right) ,
\end{multline}
where $m$ is the physical particle mass, $a$ is the scattering length in the nuclear physics convention ($a>0$ for repulsive interactions) and ${\cal I}$ and ${\cal J}$ are geometric constants given by\footnote{Huang and Yang report their result in terms of $C \equiv - \mathcal I/\pi$ and $\xi \equiv \mathcal J$. Their result contains a numerical error, reporting $C=2.37$ which differs significantly from the correct value of $C=2.84$. This was pointed out in Ref.~\cite{Beane2007}.}
\begin{eqnarray}
  \label{eq:2}
  {\cal I}&=&\lim_{\Lambda\to\infty}\sum_{\bf n\ne 0}^{|{\bf n}|\leq\Lambda} \frac{1}{|{\bf
      n}|^2} -4\pi\Lambda = -8.91363291781 \,, \ \ \ 
{\cal J}\ =\ \sum_{\bf n\ne 0} \frac{1}{|{\bf n}|^4}  = 16.532315959 \,.
\end{eqnarray}
One important consequence of this result is that the relevant expansion parameter is $a/L$ and thus the expansion is only valid for $a \ll L$. By contrast, L\"uscher's result holds for arbitrarily large (indeed divergent) scattering lengths. 

Higher orders in the $1/L$ expansion have been calculated in a number of recent papers. At $\mathcal O(1/L^6)$ comparing different calculations is no longer straightforward, and it is useful to summarize the various calculations in some detail. [See also Table \ref{tab:tab}.]

\begin{table}
\begin{center}
\begin{tabular}{c || c | c }
\hline
 & $n=2$ & $n = 3$ \\ \hline & &  \\[-10pt] 
Beane, et.~al.~\protect\cite{Beane2007}    & $  \frac{8\pi^2 a^3 r}{m } +  \mathcal C_2 \frac{ a^4}{  m }$ & $ \frac{24\pi^2 a^3 r}{m }
+  \eta_3(\mu) +  \mathcal C_{\mathrm{log}} \frac{  a^4}{m } \log(\mu L) + \mathcal C_3 \frac{ a^4}{  m } $ \\[4pt] \hline & & \\[-10pt]
Tan \protect\cite{Tan2007}   & NA & $ \frac{36\pi^2 a^3 r}{m }
+  \frac{D}{m} +  \mathcal C_{\mathrm{log}} \frac{a^4}{m } \log(L/\vert a \vert) + \mathcal D_3 \frac{ a^4}{  m }   $  \\[4pt] \hline & & \\[-10pt]
Exp. L\"uscher \protect\cite{ourpt} & $     \frac{8\pi^2 a^3 r}{m } - \frac{4 \pi^2 a^2}{m^3} + \mathcal C_2 \frac{ a^4}{  m }$ & NA  \\[4pt] \hline & & \\[-10pt]
$\lambda \phi^4$ \protect\cite{ourpt} & $- \frac{3 \lambda^2}{256 m^5} + \frac{\lambda^3}{768 \pi^2 m^5} + \mathcal O(\lambda^4) $ &$\frac{72 \pi^2 a^3 r}{m}  - \frac{\mathcal M_{3,\mathrm{th}}}{48 m^3}$   \hspace{80pt}        \\[5pt]  & $\Big ( = \frac{8\pi^2 a^3 r}{m } - \frac{4 \pi^2 a^2}{m^3} + \mathcal O(a^4) \Big )$  &        \hspace{40pt}   $ +\frac{768 a^3 \pi^3   C_3}{m^2} + \frac{36 \pi^2 a^2}{m^3} + \mathcal O(a^4)$ \\[4pt] \hline
\end{tabular}
\end{center}
\caption{Summary of results for the coefficient of the $1/L^6$ term appearing in the two- and three-particle ground state energies. Here we have introduced the geometric constants $\mathcal C_2 = 104.45105$, $\mathcal C_3=2608.5851$, $\mathcal D_3=3613.5625$ and $\mathcal C_{\mathrm{log}}=64 \pi (3 \sqrt{3} - 4 \pi)$. For $\lambda \phi^4$ theory we have given the result in terms of $\lambda$, defined as minus the threshold two-to-two scattering amplitude, $\lambda \equiv 32 \pi m  a$. We have also rewritten this in terms of $a$ and $r$ to show agreement with the expansion of L\"uscher's result. [Note that $a$ and $a^2 r$ both start at $\mathcal O(\lambda)$.] In the case of two particles, a term proportional to $a^2/m^3$ arises in the expansion of L\"uscher's result and the $\lambda \phi^4$ calculation. This term is absent in the non-relativistic calculation. In the three-particle case the result of Ref.~\cite{Beane2007} is scheme-dependent when expressed in terms of $\eta_3(\mu)$ and the present form, in particular the value of $\mathcal C_3$, is determined in the minimal subtraction $(\mathrm{MS})$ scheme. The $\lambda \phi^4$-theory three-particle result is expressed in terms of the subtracted three-particle amplitude, $\mathcal M_{3,\mathrm{th}}$, and the constant $C_3$, both explained in the text.}

\vspace{-10pt}

\label{tab:tab}
\end{table}

In Ref.~\cite{Beane2007}, Beane, Detmold and Savage calculate the $1/L^6$ contribution to the ground state using non-relativistic quantum mechanics with three-body interactions described by a delta-function potential. The result applies for any number of particles, $n$. For $n \geq 3$ the coefficient of the three-body potential, denoted $\eta_3$, enters the energy shift at $\mathcal O(1/L^6)$. 

Tan performed a closely related calculation in the same year, working through $\mathcal O(1/L^7)$ but restricting attention to three particles \cite{Tan2007}. Tan's calculation is based on an asymptotic expansion of the three-particle wave function at large interparticle distances. Using this expansion he defines a three-particle hypervolume, $D$, which is the analog of the scattering length. It is this quantity that enters his calculation at $\mathcal O(1/L^6)$, in a manner similar to $\eta_3$ in Ref.~\cite{Beane2007}.

As a result of the two different parameters, $\eta_3$ and $D$, comparison of the two studies for $n=3$ does not provide a check on the $1/L^6$ term. One does however find agreement for the logarithmic volume dependence that arises at this order [see Table \ref{tab:tab}]. Going further, in Ref.~\cite{Detmold:2008gh}, Detmold and Savage pushed the calculation for $n$ bosons through $\mathcal O(1/L^7)$. No new parameters arise at this order, so if $D$ and $\eta_3$ are constrained via the $1/L^6$ term, then unambiguous comparison is possible. For $n=3$ the authors find full agreement with the earlier work by Tan.

The calculations of Refs.~\cite{Beane2007,Tan2007,Detmold:2008gh} all use non-relativistic techniques. One can directly compare to a relativistic result in the two-particle case by expanding L\"uscher's quantization condition order by order in $1/L$. This has been done recently by the present author, together with Steve Sharpe, in Ref.~\cite{ourpt}[see again Table \ref{tab:tab}]. We find a term scaling as $a^2/m^3$, which does not appear in the two-particle non-relativistic results~\cite{Beane2007,Detmold:2008gh}. The expansion of L\"uscher's result is performed to check the main focus of Ref.~\cite{ourpt}: a calculation of the energy shift for two and three particles in relativistic $\lambda \phi^4$ theory. The $\lambda \phi^4$ theory result reproduces L\"uscher's expansion in the two-particle case, and gives a prediction for the relativistic energy shift in the case of three particles. 

The main motivation of the $\lambda \phi^4$ calculation is to provide a cross check for the relativistic three-particle quantization condition discussed in Sec.~\ref{sec:rel} below. We find complete agreement between the perturbative prediction and the expansion of the general form~\cite{ourexpqc}. In expanding the relativistic three-particle result, the three-body interaction arises at $\mathcal O(1/L^6)$ in the form of a subtracted, threshold three-to-three scattering amplitude, denoted $\mathcal M_{3,\mathrm{th}}$. 

\begin{figure}
\begin{center}
\vspace{-10pt}
\includegraphics[scale=0.38]{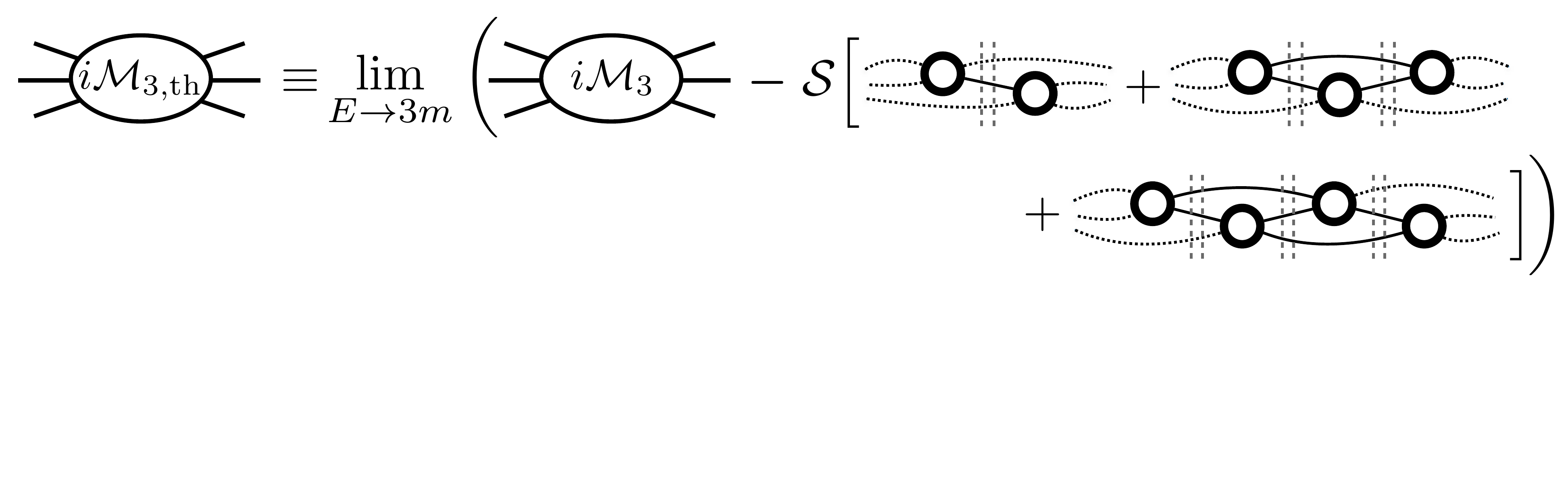}
\vspace{-75pt}
\end{center}

\caption{We define a threshold scattering amplitude by subtracting the singular parts of divergent diagrams before sending the energy to $3m$. The rings denote the two-to-two scattering amplitude near threshold. The vertical dashed lines indicate that a simple pole is used in place of the fully dressed propagator. Here $S$ indicates that the external legs of the subtracted diagrams are symmetrized in incoming and outgoing momenta. See Ref.~\cite{ourpt} for details.}
\label{fig:Mthr}
\end{figure}

The usual three-to-three amplitude diverges at threshold due to pairwise scattering diagrams, and only by subtracting off these divergent terms can one define a finite quantity. In the case of three identical particles, three such diagrams contain divergences which must be removed [see Fig.~\ref{fig:Mthr}]. The subtraction depends on a cutoff in the loop diagrams and this leads to a cutoff-dependent constant $C_3$ appearing in the $1/L^6$ result. As is explained in detail in Ref.~\cite{ourpt}, all cutoff dependence cancels between the term containing $\Mthr$ and that containing $C_3$.

We close this section by highlighting a recent result by Detmold and Flynn, presented in Ref.~\cite{Detmold:2014fpa}. In this work the authors consider matrix elements of a weak current $J$ between $n$-particle finite-volume states. Expanding the finite-volume matrix elements in powers of $1/L$, the authors give an expression in terms of constants which parametrize the current, together with the two-particle scattering length and various geometric constants. In contrast to the ground state energies discussed above, the $L \rightarrow \infty$ limit of the current contains non-trivial physical information. In particular the authors find a $\mathcal O(L^0)$ term which depends only on the one-body current in their expansion. This receives finite-volume correction which start at $\mathcal O(1/L^2)$ and the authors calculate three nontrivial orders, $1/L^2,1/L^3,1/L^4$.

\section{Non-relativistic, nonperturbative analyses}

\label{sec:nonrel}

In this section we review various nonperturbative analyses that investigate the properties of three-particle finite-volume states. We restrict attention to three recent studies, all based in non-relativistic quantum mechanics.

We start with the work of Polejaeva and Rusetsky presented in Ref.~\cite{Polejaeva}. In this extensive study, the Fadeev equations are used to analyze the three-boson spectrum. After explaining the equations in detail the authors show how to recast these in order to describe the system in a finite-volume. Various technical details arise, including a proof that poles associated with the free spectrum cancel in the finite-volume correlator. The authors also identify the appearance of finite-volume effects associated with the two-particle cusp and show how these can be accommodated. Such cusps are also an issue in the relativistic analysis described in Sec.~\ref{sec:rel} below.

The authors then give a three-particle analog of the L\"uscher equation by making a conjecture for the finite-volume Fadeev equations and then showing that this produces the correct finite-volume diagrammatic series. A key consequence of the result is the proof that, up to exponentially suppressed effects, the three-particle spectrum is determined by the on-shell scattering amplitudes. However, the resulting equations are complicated and the authors stress that future work is needed to show how the amplitudes can be extracted.

We now turn to the analyses of Brice\~no and Davoudi in Ref.~\cite{raulthree}. In this work the authors study the spectrum for three identical bosons by making use of an auxiliary s-wave dimer field. This field non-perturbatively sums all two-to-two diagrams, thereby simplifying the set which must be summed to define a finite-volume three-particle correlator. The method requires truncation of the effective range expansion and also projection to the s-wave in the two-particle sector. 

Within these approximations the authors identify a closed expression for the three-particle finite-volume correlator. The result is a geometric series built from alternating insertions of the finite-volume dimer with a three-body kernel. Summing this series, the authors identify a quantization condition that depends on an intermediate three-body quantity in which finite-volume effects persist. The authors then relate this via an integral equation to the scattering amplitude. 

As a check on the formalism, the limit in which two of the three particles are bound is considered. Here the authors recover the s-wave L\"uscher formalism with an exponentially suppressed correction, $e^{- \gamma L}$ where $\gamma$ is the binding momentum. This can be used to estimate the finite-size effects that arise in using the L\"uscher formalism to extract scattering of a particle off a bound state.

We close this section by describing important recent work by Mei\ss ner, R\'ios, and Rusetsky in Ref.~\cite{Meissner:2014dea}. In this article, the authors study the finite-volume shift to a three-body bound state. Considering a shallow bound state and working in the unitary limit ($a \rightarrow \infty$) the authors find
\begin{equation}
\label{eq:akakibs}
\Delta E = c \frac{\kappa^2}{m} \frac{1}{(\kappa L)^{3/2}} \vert A \vert^2 \exp \left( - \frac{2}{\sqrt{3}} \kappa L  \right ) + \cdots \,,
\end{equation}
where the ellipsis stands for subleading terms in $L$, both exponentially and power suppressed. Here $c=-87.866$ is a known numerical constant and $\kappa$ is the binding momentum, related to the binding energy via $E_B = \kappa^2/m$. The authors describe $\vert A \vert^2$ as the three-body analog of the asymptotic normalization coefficient of the bound state wave function. This quantity is expected to be near unity if long range effects dominate in the creation of the bound state. Eq.~(\ref{eq:akakibs}) represents an elegant and concrete prediction and reproducing this result with the general formalism discussed in the next section would be an interesting and non-trivial check on the latter.

\section{Relativistic QFT in a finite-volume}

\label{sec:rel}

We now turn to analysis of a three-particle system based in relativistic quantum field theory. This work, by the present author together with Steve Sharpe, is presented in Refs.~\cite{us} and \cite{KtoM}. 

\subsection{Set up}

In these references we consider identical scalar particles with physical mass $m$ contained in a periodic cubic spatial volume with extent $L$. For a given total momentum in the finite-volume frame, $\vec P \in (2 \pi/L) \mathbb Z^3$, we relate scattering observables to the discrete finite-volume energy spectrum of three-particle states. In the derivation we drop exponentially suppressed corrections of the form $e^{-mL}$, but keep all terms which scale as some power of $1/L$. We use $E$ to denote the three-particle energy in the moving frame and $E^*$ the energy in the center of mass (CM) frame ($E^* = \sqrt{E^2 - \vec P^2}$).

In the derivation of Refs.~\cite{us, KtoM}, particle interactions are described by a Lagrangian with a $\mathbb Z_2$ symmetry that prevents even-odd coupling, i.e.~only contains even powers of the single-particle interpolating field. This is a technical simplification and has the nice feature that we can meaningfully separate the odd- and even-number states. We study the spectrum in the odd-particle sector, for CM energies which lie above the one-particle pole and below the five-particle production threshold, $m<E^*<5m$.

One additional assumption is needed for the formalism to hold. We must require that the two-particle K matrix, defined in Eq.~(\ref{eq:Ksdef}) below, has no poles in the range of allowed two-particle energies. Practically this means that, if the system contains a two-particle resonance with mass $m_R$ then the kinematic range of validity is further restricted: 
$
m < E^* < \mathrm{min}[m+m_R, \ 5m] 
$. 
This ensures that the maximum energy in a two-particle subsystem, $E^*-m$, is below the resonance pole.

Before going into detail about the finite-volume analysis, it is instructive to examine the infinite-volume quantities which are expected to appear in the study. By analogy to L\"uscher's work together with that of the previous sections, we expect that, for three-particle states, both the two-to-two amplitude $\mathcal M_2$ and the three-to-three amplitude $\mathcal M_3$ should play a role. 

The two-to-two amplitude, $\mathcal M_2$, is a function of the CM energy, $E_2^*$, as well as the direction of back-to-back momenta in the CM frame. We find it convenient to separately label an incoming direction, $\hat a^*$, and an outgoing direction, $\hat a'^*$. For fixed energy, $\mathcal M_2$ is known to be a smooth function of these angles. This motivates a decomposition in partial waves, leading to coefficients given by $\mathcal M_{\ell} (E^*)$ for $\ell = 0, 1, 2, \cdots$. These are often expressed in terms of scattering phase shifts.

\begin{figure}
\begin{center}
\vspace{-20pt}
\includegraphics[scale=0.38]{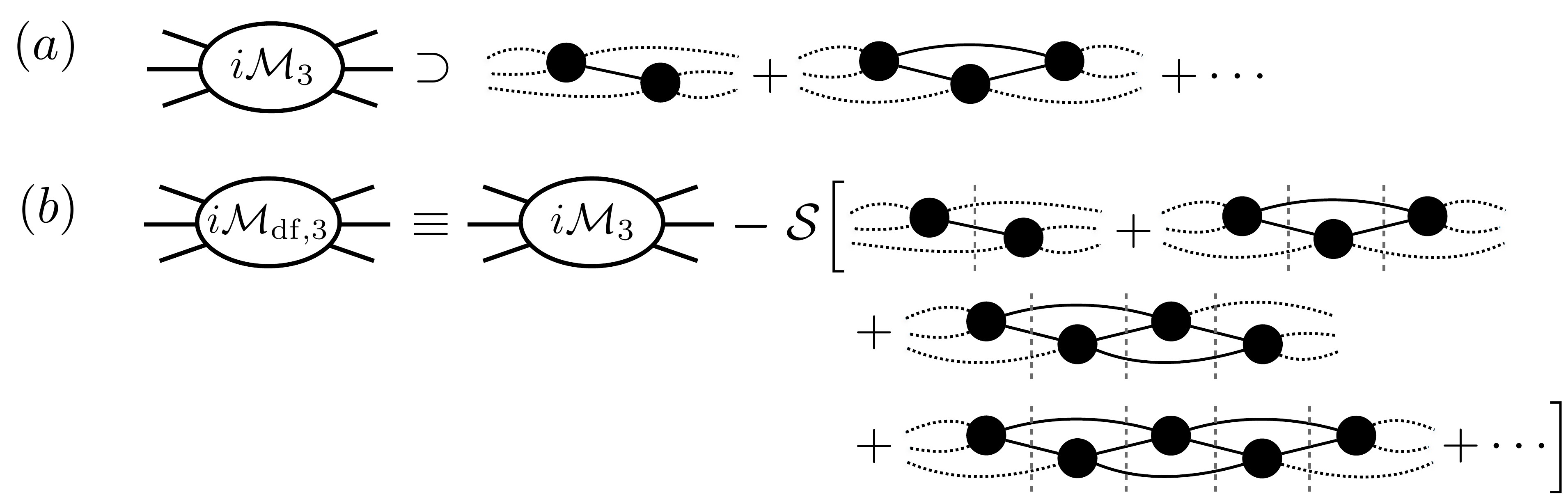}
\caption{(a) Example of the diagrams contributing singularities to $\mathcal M_3$. Here filled circles represent two-to-two scattering amplitudes, connected by internal propagators which are generally off-shell. For certain external momenta the internal propagators go on-shell, resulting in divergences in $\mathcal M_3$. Subtracting these gives (b) the divergence-free three-to-three
  amplitude, \(\Mdf\). In the subtracted term, filled circles
  represent on-shell two-to-two scattering amplitudes \(\M\). Dashed
  cuts stand for simple kinematic factors that appear between adjacent $\M$. These factors have the requisite poles so that the subtracted terms cancel the singularities in $\Mth$.
 The \( S\) outside the square
  brackets indicates that the subtracted terms are symmetrized.}
\label{fig:Mdfdef}
\end{center}

\vspace{-15pt}

\end{figure}

Naturally, $\mathcal M_3$ depends on additional degrees of freedom. For fixed energy and momentum, $E, \vec P$, we denote the remaining functional dependence by $\vec k, \hat a^*$ for the instate and $\vec p, \hat a'^*$ for the outstate. In each case the three-vector ($\vec k$, $\vec p$) is the momentum of one of the three particles in the finite-volume frame. The starred unit vector ($\hat a^*$, $\hat a'^*$) is the direction of back-to-back momentum for the remaining two in their CM frame. In contrast to $\mathcal M_2$, the three-to-three scattering amplitude is not a smooth function of these coordinates. The reason is that pairwise scattering diagrams lead to divergences, associated with long lived intermediate states (see Fig.~\ref{fig:Mdfdef}(a) as well as Refs.~\cite{Taylor:1977A,Taylor:1977B, Brayshaw:1969ab}). Such divergences appear for all $E^*>3m$, whenever $\vec k, \hat a^*, \vec p, \hat a'^*$ coincide with the momenta of classical pairwise scattering events. This leads us to define a divergence-free three-to-three scattering amplitude, $\Mdf$ [see Fig.~\ref{fig:Mdfdef}(b)]. The precise definition of $\Mdf$ is given in Eq.~(87) of Ref.~\cite{KtoM}. We stress here that the difference between $\Mdf$ and $\mathcal M_3$ only depends on on-shell values of $\M$ and that, unlike $\Mth$, $\Mdf$ is a smooth function that can be decomposed in generalized harmonics with the lowest modes dominating at low energies.

\subsection{Skeleton expansion}

\begin{figure}
\begin{center}
\includegraphics[scale=0.38]{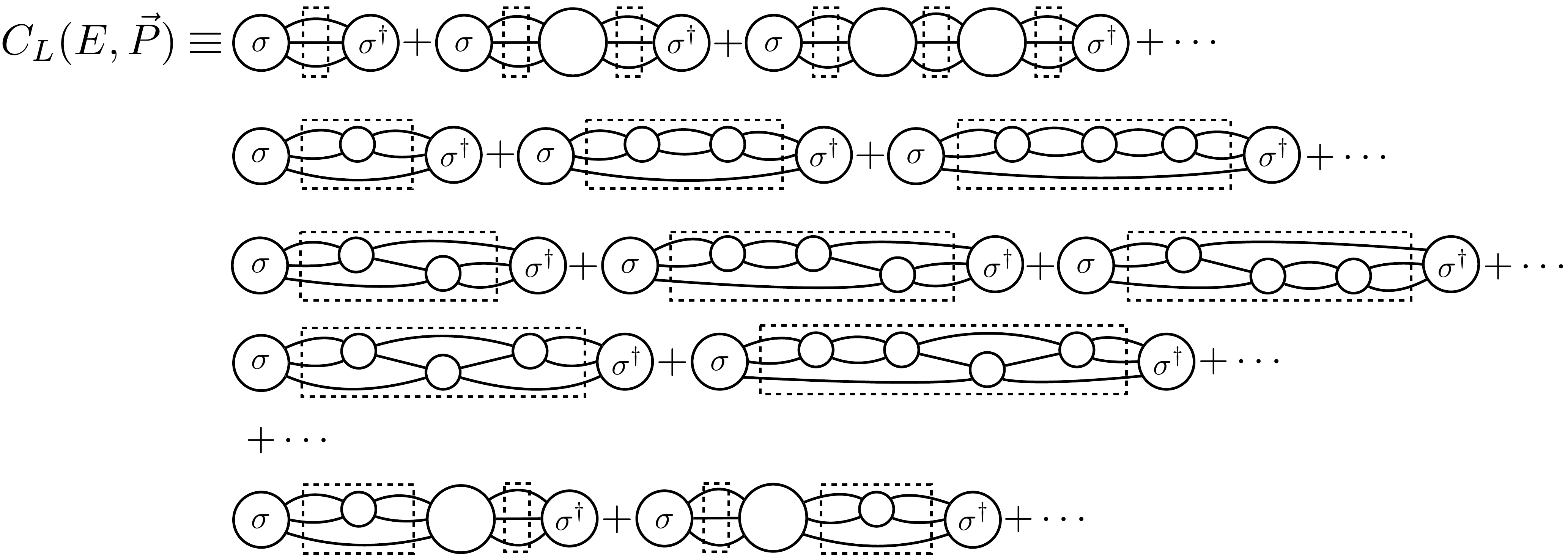}
\vspace{-25pt}
\end{center}
\caption{Skeleton expansion for the finite-volume correlator. 
  Outermost blobs in all diagrams represent functions of momentum 
  that are determined by the interpolating operators 
  $\sigma$ and $\sigma^\dagger$. 
  Insertions between these functions having four legs represent two-to-two
  Bethe-Salpeter kernels, \(i B_{2}\), while insertions with six legs
  represent the analogous three-to-three kernels, \(i B_{3}\). 
  Lines connecting kernels and \(\sigma\)-functions represent
  fully-dressed propagators. The kernels and
  dressed propagators can be replaced by their infinite-volume 
  counterparts (in which internal loop momenta are integrated).
  However, the spatial momenta flowing along the propagators that are
  shown explicitly, and that lie
  within the dashed rectangles, are summed rather than integrated.}
  \label{fig:CL}
\end{figure}

Returning to the finite-volume system, the derivation presented in Ref.~\cite{us} is organized in analogy to the two-particle field theoretic study of Ref.~\cite{KSS}. We introduce a finite-volume correlator, 
\begin{equation}
C_L(E, \vec P) \equiv \int_{L} d^4 x 
e^{i(E x^0-\vec P \cdot \vec x)} 
\langle 0 \vert \mathrm{T} \sigma(x) \sigma^\dagger(0) \vert 0
\rangle \,,
\end{equation}
whose poles in $E$ give the finite-volume energy spectrum at fixed $\{L,\vec P\}$. Here $\sigma$ and $\sigma^\dagger$ are odd-particle interpolating fields. We express this correlator using a skeleton expansion built from two types of Bethe-Salpeter kernels, connected by fully dressed propagators and with endcaps determined by the interpolating functions. This skeleton expansion is summarized in Fig.~\ref{fig:CL}.

The choice to express $C_L(E, \vec P)$ as indicated is motivated by the observation that, for the kinematics considered, the two kernels, the fully dressed propagators and the interpolating functions all have exponentially suppressed volume dependence. That is, the difference between the finite- and infinite-volume versions of these quantities scales as $e^{-mL}$ for large $L$. As mentioned above, we take these to be negligible and thus work with the infinite-volume versions of kernels, propagators, and interpolators throughout. Thus, the only $L$ dependence in $C_L$, as expressed in Fig.~\ref{fig:CL}, is due to the sums over momenta in the explicitly displayed loops. 

The remaining task of the derivation is to rewrite these sums as integrals plus finite-volume residues, to sum all integrals into infinite-volume quantities and to thus reach an expression relating the finite-volume spectrum to infinite-volume scattering observables. In Ref.~\cite{us} the separation of finite- and infinite-volume quantities as well as the summation into a closed expression is achieved. However the infinite-volume three-particle quantity that appears is nonstandard due to technical issues that arise in the derivation. The nonstandard quantity is referred to as the divergence-free three-to-three K matrix and is denoted by $\Kdf$. In Ref.~\cite{KtoM} this nonstandard quantity is related, via purely infinite-volume integral equations, to the divergence-free three-to-three amplitude $\Mdf$ and to the standard amplitude $\Mth$.

\subsection{Sketch of the two-particle derivation}

\vspace{0pt}

\begin{figure}
\begin{center}
\includegraphics[scale=0.5]{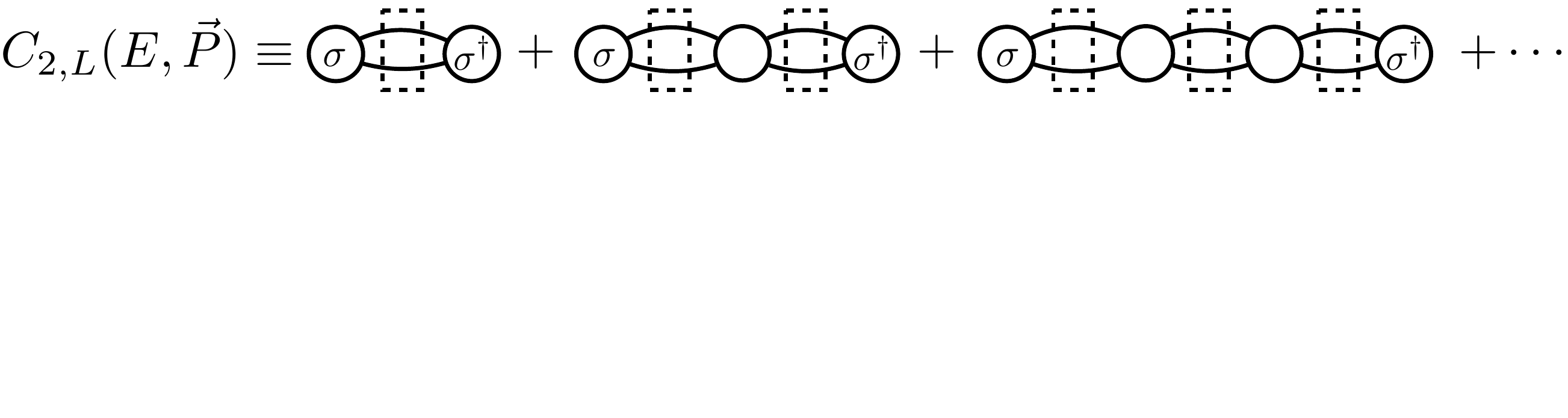}
\vspace{-85pt}
\end{center}
\caption{Skeleton expansion for the two-particle correlator. Outermost blobs represent interpolators, those in between the kernel, $i B_2$, and all lines represent fully dressed propagators. As in the three-particle case, the finite-volume dependence of interpolators, kernels and propagators is exponentially suppressed, and only the $L$ dependence due to the sums in two-particle loops is included.}
  \label{fig:CL2}
\end{figure}

To give some indication of the derivation in Ref.~\cite{us}, here we describe the analogous steps in the case of two particles. Following Ref.~\cite{KSS}, we begin with the skeleton expansion for a two-particle correlator with CM energy satisfying $0 < E_2^* < 4m$ [see Fig.~\ref{fig:CL2}]
\begin{equation}
C_{2,L}(E, \vec P) = \sum_{n=0}^\infty \sigma i S \left[ i B_2 i S \right ]^n \sigma^\dagger \,,
\end{equation}
where $S$ stands for sum and we have introduced the shorthand
\begin{equation}
iB_2 i S i B_2 = \frac{1}{L^3} \sum_{\vec k} \int \frac{d k^0}{2 \pi} i B_2 (p', k)  \Delta(k)  \Delta(P-k)     i B_2(k,p) \,,
\end{equation}
where $\Delta(k)$ is the fully dressed single-particle propagator, renormalized to have residue $i$ at the single particle pole. Here we have also introduced $i B_2(k,p)$ as the two-to-two Bethe-Salpeter kernel with incoming momenta $p$ and $P-p$ and outgoing momenta $k$ and $P-k$. Next define an analogous, infinite-volume version
\begin{equation}
iB_2 i I^{\mathrm{PP}} i B_2 = \mathrm{PP} \int \frac{d^4 k}{(2 \pi)^4} i B_2 (p', k)  \Delta(k)  \Delta(P-k)     i B_2(k,p) \,.
\end{equation}
Here ``PP'' stands for pole prescription. This represents the fact that we have freedom to choose the prescription used to define the integral over $\vec k$. 

Finally, introducing the sum-integral-difference $F^{\mathrm{PP}}$ by $iB_2 i F^{\mathrm{PP}} i B_2 \equiv iB_2 i S i B_2 - iB_2 i I^{\mathrm{PP}} i B_2$, we can rewrite $C_{2,L}(E, \vec P) $ as follows
\begin{align}
\label{eq:subbed}
C_{2,L}(E, \vec P) & = \sum_{n=0}^\infty \sigma ( i I^{\mathrm{PP}} + i F^{\mathrm{PP}}  ) \left[ i B_2 ( i I^{\mathrm{PP}} + i F^{\mathrm{PP}}  ) \right ]^n \sigma^\dagger \,, \\
& = C_\infty^{(2), \mathrm{PP}}  + i A^{' \mathrm{PP}} F^{\mathrm{PP}} \frac{1}{1 + \mathcal M^{\mathrm{PP}} F^{\mathrm{PP}}}  A^{ \mathrm{PP}} \,,
\label{eq:summed}
\end{align}
where
\begin{gather}
\label{eq:infdefs1}
 C_{2,\infty}^{\mathrm{PP}}  \equiv \sum_{n=0}^\infty \sigma i I^{\mathrm{PP}} \left[ i B_2 i I^{\mathrm{PP}} \right ]^n \sigma^\dagger\,, \ \ \ \ \  \ \ A^{' \mathrm{PP}} \equiv \sum_{n=0}^\infty \sigma  \left[ i I^{\mathrm{PP}} i B_2  \right ]^n  \,, \\
  A^{\mathrm{PP}} \equiv \sum_{n=0}^\infty   \left[ i B_2 i I^{\mathrm{PP}} \right ]^n \sigma^\dagger \,, \ \ \ \ \ \ \   i \mathcal M^{  \mathrm{PP}} \equiv \sum_{n=0}^\infty  \left[ i B_2 i I^{\mathrm{PP}} \right ]^n i B_2 \,.
\label{eq:infdefs2}
\end{gather}
To reach Eq.~(\ref{eq:summed}) from Eq.~(\ref{eq:subbed}) we have simply rearranged the terms according to the number of $F^{\mathrm{PP}}$ insertions and grouped the expressions between insertions into various infinite-volume quantities, defined in Eqs.~(\ref{eq:infdefs1}, \ref{eq:infdefs2}).

We now give three important observations about the result, Eq.~(\ref{eq:summed}). First note that the finite-volume correlator, $C_{2,L}$, cannot depend on the pole-prescription used to evaluate integrals over spatial momentum, $d^3 k$, that appear on the right-hand side. We introduced this ``PP'' dependence in the integral $I^{\mathrm{PP}}$ and the difference $F^{\mathrm{PP}}$, and we know that it cancels in the sum of these two quantities. This cancellation is not manifest in Eq.~(\ref{eq:summed}), but must still hold. 

Second, as is discussed extensively in Refs.~\cite{KSS,us}, the four-vectors within $iB_2$, which are summed in $S$ and integrated in $I^{\mathrm{PP}}$ are projected on-shell in the difference $F^{\mathrm{PP}}$. Boosting to the CM frame we reach back-to-back four-vectors $(\sqrt{q^{*2} + m^2}, q^* \hat k^*)$ and $(\sqrt{q^{*2} + m^2}, - q^* \hat k^*)$, where $q^*$ is given by $E^* = 2 \sqrt{q^{*2} + m^2}$. It follows that $\hat k^*$ is the only remaining degree of freedom. This motivates us to decompose  $A^{' \mathrm{PP}}, A^{\mathrm{PP}}$ and $\mathcal M^{  \mathrm{PP}}$ in spherical harmonics, and view the second term in Eq.~(\ref{eq:summed}) as a matrix product on the resulting index space. In particular $A^{' \mathrm{PP}}_{\ell, m}$ should be viewed as a row, $F^{  \mathrm{PP}}_{\ell' m'; \ell m}$ and $\mathcal M^{  \mathrm{PP}}_{\ell' m'; \ell m}$ as matrices, and $A^{ \mathrm{PP}}_{\ell, m}$ as a column. See Refs.~\cite{KSS,us} for details.

Third and finally, we return to the main motivation for considering $C_{2,L}(E, \vec P)$, the fact that poles in $E$ give the finite-volume spectrum of the theory. These poles arise from the only finite-volume dependent factor, that appearing between $A'^{\mathrm{PP}}_{\ell, m}$ and $A^{ \mathrm{PP}}_{\ell, m}$. Various equivalent ways of expressing the divergence are possible, for example
\begin{equation}
\label{eq:qctwo}
\mathrm{det}_{\ell, m}\big [ (\mathcal M^{\mathrm{PP}})^{-1} + F^{\mathrm{PP}} \big ] = 0 \,.
\end{equation}
This is the two-particle quantization condition due to L\"uscher, expressed here in a pole-prescription-agnostic way. If we choose PP to be the $i \epsilon$ prescription then $\mathcal M^{\mathrm{PP}}$ becomes the standard scattering amplitude, $\M$, while if we choose it to be the principal-value prescription, then $\mathcal M^{\mathrm{PP}}$ becomes the K matrix, $\K$. Indeed it is well known that $(\K)^{-1} = \mathrm{Re}[(\M)^{-1}]$, and one can also show that $F^{\mathrm{pv}} = \mathrm{Re}[F^{i \epsilon}]$ and that the imaginary parts of $(\M)^{-1}$ and $F^{i \epsilon}$ exactly cancel. This makes the equivalence of the two pole prescriptions manifest.

\subsection{Unitary cusps and $\K$}

\begin{figure}
\begin{center}
\includegraphics[scale=0.25]{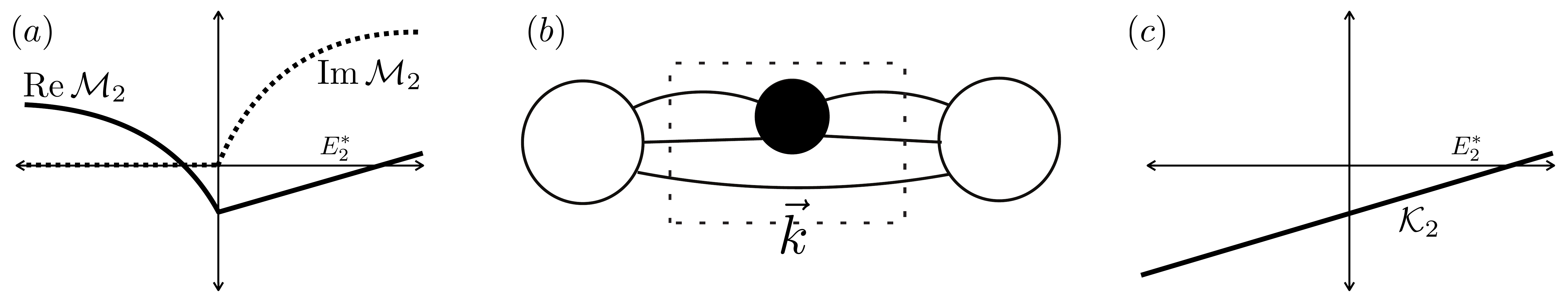}
\vspace{-10pt}
\end{center}
\caption{(a) Sketch of $\M$ as a function of energy, with a cusp at $E_2^*=2m$ (b) Example of a diagram in which the sum crosses the cusp (c) $\K$, in contrast to $\M$, is smooth at two-particle threshold.}
  \label{fig:cusp}
\end{figure}

In the case of the three-particle quantization condition, technical issues lead us to use a non-standard pole prescription in the separation of sums into integrals and sum-integral differences. The complication is that in the standard $i \epsilon$ prescription, the unitary cusp of $\M$ at two-particle production threshold [see Fig.~\ref{fig:cusp}(a)] induces finite-volume effects. This arises in diagrams such as that shown in Fig.~\ref{fig:cusp}(b). Since the total energy and momentum are fixed, the sum over the momentum of the bottom propagator varies the energy flowing through the two-to-two amplitude. This means the sum sweeps across the cusp, inducing a new kind of finite-volume effect.

Rather than explicitly quantify and sum all cusp finite-volume effects, in Ref.~\cite{us} we use a new pole prescription that replaces the two-to-two scattering amplitude with a modified K matrix. The latter has no cusp at two-particle threshold, and is indeed a smooth function for all two-particle energies $E_2^*$ [see Fig.~\ref{fig:cusp}(c)]. This relies on the requirement that the modified K matrix contains no poles, discussed above. The idea to use a different pole prescription to remove the cusp was first introduced in the non relativistic analysis Ref.~\cite{Polejaeva}. This has the consequence that the three-to-three quantity that emerges is not the standard three-to-three scattering amplitude.

To give a more complete indication of the modified pole prescription, we now completely define the s-wave component of $\K$
\begin{equation}
\label{eq:Ksdef}
\K^{s}(E_{2,k}^*)^{-1} \equiv \M^{s}(E_2^*)^{-1} - H(\vec k) \tilde \rho(E_{2,k}^*) \,.
\end{equation} 
Here $\M^{s}$ is the standard s-wave two-to-two scattering amplitude. We have defined both $\K^s$ and $\M^s$ as functions of the two-particle CM energy, $E_{2,k}^*$. In the context of the three-particle quantization condition, this quantity arises when two of the three particles scatter while the third spectates [see Fig.~\ref{fig:cusp}(b)]. The subscript $k$ on $E_{2,k}^*$ indicates that the resulting two-particle energy is completely specified by the spectator momentum $\vec k$ together with the total energy and momentum,
$
E_{2,k}^* \equiv \sqrt{(E - \omega_k)^2 - (\vec P - \vec k)^2} \,,
$ 
where $\omega_k \equiv \sqrt{\vec k^2 + m^2}$.  

The factors in the second term of Eq.~(\ref{eq:Ksdef}) are defined as\footnote{Other definitions of $H(\vec k)$ are possible but we focus on a particular implementation for concreteness. See the discussion of Ref.~\cite{us} for full details.}
\begin{align}
\label{eq:rhotildef}
\tilde\rho(E_2^*) &\equiv \frac{1}{16 \pi  E_2^*} \times
\begin{cases} 
-  i \sqrt{E_2^{*2}/4-m^2} \,, & (2m)^2< E_2^{*2} \,;
\\[5pt]
\left \vert \sqrt{E_2^{*2}/4-m^2} \right \vert \,, &   0< E_2^{*2} \leq (2m)^2 \,,
\end{cases} \\[10pt]
H(\vec k) & \equiv J(E_{2,k}^{*2}/[4m^2]) \,, \ \ \ \ \ J(x) \equiv
\begin{cases}
0 \,, & x \le 0 \,; 
\\ 
\exp \left( - \frac{1}{x} \exp \left [-\frac{1}{1-x} \right] \right ) \,, 
& 0<x \le 1 \,; 
\\ 
1 \,, & 1<x \,.
\end{cases}
\end{align}
To explain the nature and origin of these definitions it is useful to consider different regions of $\vec k$, leading to different values of $E^*_{2,k}$. First suppose that $E, \vec P$ and $\vec k$ are such that $2m<E_{2,k}^*$. Then $H(\vec k)=1$, leading to
\begin{equation}
\K^{s}(E_{2,k}^*)^{-1} = \M^{s}(E_2^*)^{-1} -  \tilde \rho(E_{2,k}^*) = \mathrm{Re}[\M^{s}(E_2^*)^{-1}] \,, \ \ \ \ \ \ \ \mathrm{for\ } 2m < E_{2,k}^*   \,,
\end{equation} 
where we have used the result $\tilde \rho$ perfectly cancels the imaginary part of $\M^{s}(E_2^*)^{-1}$. In other words, for $2m<E_{2,k}^*$, $\K^{s}$ is just the standard two-particle K matrix, related to the scattering phase shift via
\begin{equation}
\K^{s}(E_{2,k}^*) = \frac{16 \pi E^*_{2,k}}{q_k^*} \tan \delta(q_k^*) \,, \ \ \ \ \ \ \ \mathrm{for\ } 2m<E_{2,k}^*   \,,
\end{equation}
where we have introduced $q_k^*$ defined by $E_{2,k}^* \equiv 2 \sqrt{q_k^{*2}+m^2}$. 

The nonstandard nature of $\K^{s}(E_{2,k}^*)$ arises from its definition for $0 < E_{2,k}^* < 2m$. In this region $\mathcal M_{2}^s$ is defined via analytic continuation and $\tilde \rho$ is continued as indicated in Eq.~(\ref{eq:rhotildef}). Indeed, if the function $H(\vec k)$ were set to one in Eq.~(\ref{eq:Ksdef}), then the right-hand side would represent the standard analytic continuation leading to subthreshold $\K^{s}(E_{2,k}^*)$. The function $H(\vec k)$ modifies this by smoothly ``turning off'' the $\tilde \rho$ term as $E_{2,k}^{*}$ varies from $2m$ to $0$. In other words, the definition smoothly interpolates from the standard K matrix at two-particle threshold to the standard scattering amplitude at $E_{2,k}^*=0$. 

If $H(\vec k)$ were not included, then two problems would arise in the three-particle quantization condition. The first is related to the fact that $\tilde \rho$ appears in $F_3$, the three-particle analog of $F^{\mathrm{PP}}$ appearing in the quantization condition Eq.~(\ref{eq:threeqc}) below. As in the two-particle case, the three-particle quantization is expressed as the determinant of a matrix, and is only useful if this matrix can be truncated to a finite-dimensional subspace. The growth of $\tilde \rho$ below threshold invalidates this truncation unless $H(\vec k)$ is included. The second issue arises when one considers higher partial waves, beyond the s-wave defined above. To do so, one needs a subthreshold definition of $\hat a^*$, the direction of back-to-back momentum in the two-particle CM frame. This definition, given in Ref.~\cite{us}, is only valid for $E_{2,k}^*>0$, and the function $H(\vec k)$ ensures that it is only needed in this regime.

\subsection{Three-particle quantization condition}

The definition of $\K^s$ in the previous section can also be interpreted as the definition of a pole prescription. In particular, to evaluate the spatial-momentum integrals for two-particle loops in $\K$, one should use a principal-value prescription for $2m<E^*_{2,k}$ and for $0< E_{2,k}^*<2m$ one should continue below threshold using $H(\vec k) \tilde \rho(E_{2,k}^*)$. In Refs.~\cite{us,KtoM} we denote this as the modified principal value, or $\PV$, prescription. This removes two-particle cusps while ensuring that we do not encounter arbitrarily large sub-threshold contributions from $\tilde \rho$. 

Using this definition, we can separate all sums in $C_L$ into integrals and sum-integral differences and sum the decomposition into a closed form, analogous to Eq.~(\ref{eq:summed}) above. Identifying the poles in this expression leads to the three-particle quantization condition
\begin{equation}
\label{eq:threeqc}
\mathrm{det}_{\vec k, \ell, m} \big [ \Kdf^{-1} + F_3 \big ] = 0 \,.
\end{equation}
Here $\Kdf$ is an infinite-volume three particle quantity which is nonstandard due to the $\PV$ pole prescription. We build up the precise definition diagram by diagram in Ref.~\cite{us} and a more useful definition is given in Ref.~\cite{KtoM} where the quantity is related to $\mathcal M_3$. $F_3$, by contrast, depends on $L$ and also on the two-to-two K matrix, $\K$. It is defined in Eq.~(19) of Ref.~\cite{us}.

As one might expect, the index space of the matrices in Eq.~(\ref{eq:threeqc}) differs from the spherical harmonic indices that arise in the two-particle case. In particular, as we have described above, with fixed total energy and momentum $E, \vec P$, an on-shell three-particle state can be parametrized by the moving-frame momentum of one of the particles, $\vec k$, together with the direction of back-to-back momentum in the CM frame of the other two, $\hat a^*$. In the quantization condition, $\vec k$ is constrained to satisfy $\vec k \in (2 \pi/L) \mathbb Z^3$. Combining this with a decomposition of $\hat a^*$ in spherical harmonics gives a discrete set, $\vec k, \ell, m$. The quantities $\Kdf$ and $F_3$ are viewed as matrices on this index space, i.e.~$\Kdf = \mathcal K_{\mathrm{df,3}; k' \ell' m' ; k \ell m}$.

The matrix index $\vec k$ is truncated due to the cutoff function $H(\vec k)$. Thus, as in the two particle result, the matrices entering Eq.~(\ref{eq:threeqc}) truncate to a finite dimensional space as long as one truncates $\K$ (inside of $F_3$) and $\Kdf$ to be finite dimensional in $\ell$ space. The most extreme truncation possible is to suppose that $\Kdf$ depends only on CM total energy $E^*$. This gives
$
\Kdf(E^*)= - 1/F^{\mathrm{iso}}_3(E,\vec P, L) 
$. 
  In this truncation, which is well motivated near threshold where directional dependence is suppressed, one can determine a value of $\Kdf(E^*)$ for each finite-volume energy level. To do so one must determine $F^{\mathrm{iso}}_3$, which is defined in Eq.~(39) of Ref.~\cite{us}. This depends on $\K$ at two-particle energies ranging up to $E^*-m$.

\subsection{Relating $\Kdf$ to $\Mdf$ and $\mathcal M_3$}

\begin{figure}
\begin{center}
\vspace{-5pt}
\includegraphics[scale=0.5]{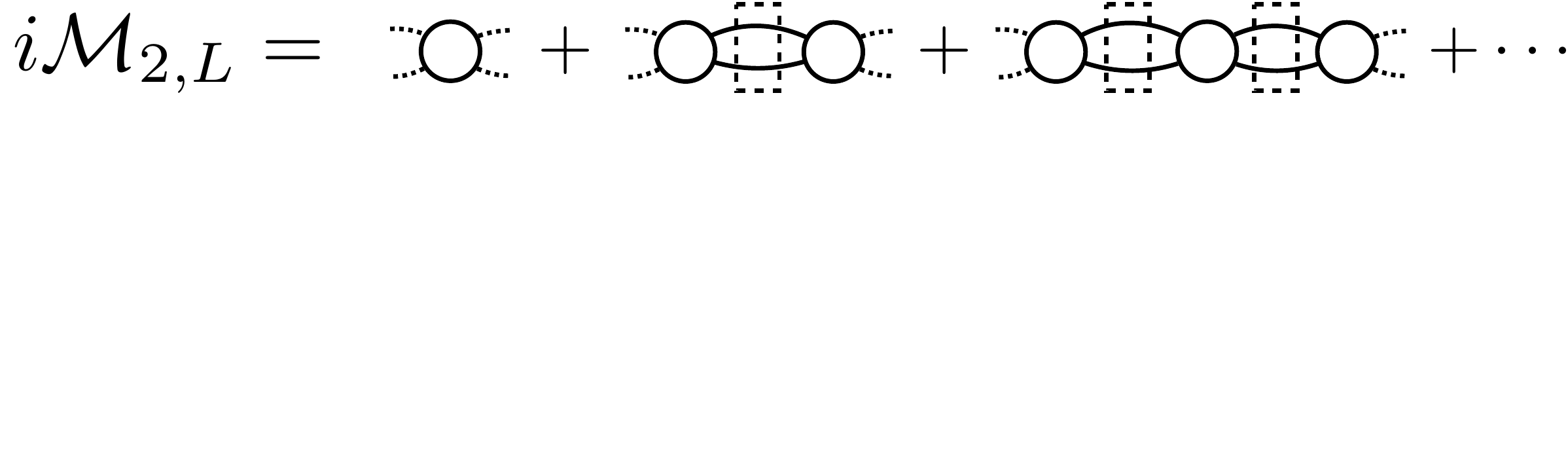}
\vspace{-90pt}
\end{center}
\caption{Alternative finite-volume two-particle correlator. This is reached by amputating the $\sigma$ and $\sigma^\dagger$ endcaps and then discarding the disconnected diagram from $C_{2,L}$ in Fig.~\protect\ref{fig:CL2}.}
  \label{fig:M2L}
\end{figure}

The precise definition of $\Kdf$ is quite complicated and our use of this non-standard infinite-volume quantity is the central drawback of Ref.~\cite{us}. This problem is resolved in Ref.~\cite{KtoM} where we give an explicit integral equation relating $\Kdf$ to the standard three-to-three scattering amplitude, $\mathcal M_3$. To give an idea of how the relation is derived it is instructive to return to the two-particle case. We begin by introducing a modification of the two-particle finite-volume correlator [see Fig.~\ref{fig:M2L}]
\begin{equation}
i \mathcal M_{2,L} \equiv \sum_{n=0}^\infty  \left[ i B_2 i S \right ]^n i B_2 = i \mathcal M^{\mathrm{PP}} \frac{1}{1 +  F^{\mathrm{PP}} \mathcal M^{\mathrm{PP}}} \,.
\end{equation}
$\mathcal M_{2,L}$ is given by choosing a specific set of interpolators in $C_{2,L}$ and also discarding disconnected diagrams. These steps do not affect the predicted spectrum and so $\mathcal M_{2,L}$ may be used in place of $C_{2,L}$ to derive Eq.~(\ref{eq:qctwo}). 

Now note that, if we can construct an infinite-volume limit such that $i S$ goes over to $i I^{i \epsilon}$, then in this limit $i \mathcal M_{2,L}$ will become the standard infinite-volume two-to-two scattering amplitude. As we describe in Ref.~\cite{us}, such a limit can be taken as follows: (1) In all summands shift all singularities into the complex plane by an amount $i \epsilon$ (2) Send $L \to \infty$ to convert the sums over the now smooth functions into integrals (3) Send $\epsilon \rightarrow 0^+$ after integration in the manner that is standard for the $i \epsilon$ pole prescription. We denote this procedure with the shorthand $\lim_{L \rightarrow \infty} \vert_{i \epsilon}$ and deduce
\begin{equation}
\label{eq:KtoMtwo}
i \mathcal M_{2} = \lim_{L \rightarrow \infty} \bigg \vert_{i \epsilon} i \mathcal M_{2,L} =  i \mathcal M^{\mathrm{PP}} \frac{1}{1 + \left[  \lim_{L \rightarrow \infty} \vert_{i \epsilon}   F^{\mathrm{PP}} \right ] \mathcal M^{\mathrm{PP}}} \,.
\end{equation}
On the left-hand side is the standard two-to-two scattering amplitude and on the right-hand side is the two-to-two quantity defined with some other pole prescription. Combining Eqs.~(\ref{eq:qctwo}) and (\ref{eq:KtoMtwo}), we have a two step procedure: first use the finite-volume spectrum to determine $\mathcal M^{\mathrm{PP}}$ and then use the infinite-volume conversion to deduce $\mathcal M_2$. In the two-particle case this is quite unnecessary because the two steps can be trivially combined into the standard L\"uscher expression. 

In the three-particle case the relations are significantly more complicated and the separation into two parts is more difficult to remove. In Ref.~\cite{KtoM} we carefully repeat the argument above in the three-particle sector. This requires first defining $\mathcal M_{3,L}$, a modified finite-volume correlator that becomes the three-to-three amplitude in the limit $\lim_{L \rightarrow \infty} \vert_{i \epsilon}$. Next we use the results for $C_L$ to express $\mathcal M_{3,L}$ in terms of $\Kdf$ as well as $\K$ and various finite-volume functions. Finally we take the infinite-volume limit to reach an integral equation relating $\Kdf$ to $\Mdf$ and to $\mathcal M_3$. 

\section{Conclusion}

\label{sec:conc}

A variety of techniques have been used to relate the three-particle finite-volume spectrum with scattering observables. In the case of $1/L$ expansions, multiple results are available and, while there are no inconsistencies in the three-particle sector, a check of the $1/L^6$ term is only possible if the various three-particle interactions can be related independently. Similarly for non-perturbative, non-relativistic studies: results are available which prove certain key properties of the spectrum, and checks between the various approaches would be instructive. 

Finally, the relativistic study of the previous section should in principal reproduce all of the results that precede it. The $1/L$ expansion is nearly complete and further checks, including the three-particle binding energy, are underway. We also aim in future work to demonstrate the practical utility of this result in extracting observables from the spectrum. This will require extending the formalism to accommodate all possible three-hadron states, in particular including non-identical, non-degenerate particles, two-particle resonances, two-to-three couplings and spin.

{\em I thank Steve Sharpe for many useful discussions and for helpful comments on the manuscript.}

\bibliographystyle{JHEP-2-notitle}
\bibliography{ref}

\end{document}